\documentclass[conference]{IEEEtran}

\usepackage[letterpaper, left=1in, right=1in, bottom=1in, top=0.75in]{geometry}
\usepackage{cite}
\usepackage[font=scriptsize]{caption}
\usepackage{subcaption}
\usepackage{algorithm}
\usepackage{algpseudocode}
\makeatletter
\def\BState{\State\hskip-\ALG@thistlm}
\makeatother

\usepackage{graphicx}
\usepackage{amsmath}
\usepackage[hidelinks]{hyperref}
\usepackage{amssymb}
\usepackage{amsthm}

\ifCLASSINFOpdf

\else

\fi
\begin{document}
\title{ Reinforcement Learning-Based Trajectory Design for the Aerial Base Stations }
\author{    \IEEEauthorblockN{Behzad Khamidehi and Elvino S. Sousa}
    \IEEEauthorblockA{
    	Department of Electrical and Computer Engineering, University of Toronto, Canada}
    Emails: b.khamidehi@mail.utoronto.ca and es.sousa@utoronto.ca}
\maketitle

\begin{abstract}
In this paper, the trajectory optimization problem for a multi-aerial base station (ABS) communication network is investigated. The objective is to find the trajectory of the ABSs so that the sum-rate of the users served by each ABS is maximized. To reach this goal, along with the optimal trajectory design, optimal power and sub-channel allocation is also of great importance to support the users with the highest possible data rates. To solve this complicated problem, we divide it into two sub-problems: ABS trajectory optimization sub-problem, and joint power and sub-channel assignment sub-problem. Then, based on the Q-learning method, we develop a distributed algorithm which solves these sub-problems efficiently, and does not need significant amount of information exchange between the ABSs and the core network. Simulation results show that although Q-learning is a model-free reinforcement learning technique, it has a remarkable capability to train the ABSs to optimize their trajectories based on the received reward signals, which carry decent information from the topology of the network. 
\end{abstract}

% no keywords
% For peer review papers, you can put extra information on the cover
% page as needed:
% \ifCLASSOPTIONpeerreview
% \begin{center} \bfseries EDICS Category: 3-BBND \end{center}
% \fi
%
% For peerreview papers, this IEEEtran command inserts a page break and
% creates the second title. It will be ignored for other modes.
\IEEEpeerreviewmaketitle

\section{Introduction}
\IEEEPARstart{S}{upporting} ceaselessly increasing number of mobile devices and their high data rate demands are among the most critical concerns of the wireless networks. Unfortunately, the currently utilized base stations (BSs) are not able to completely satisfy these requirements due to their static nature \cite{Halim_frontier}. In particular, a long distance between mobile users and these static BSs causes a low quality for the communication link, and leads to poor coverage. To resolve this issue, finding technologies allowing the BSs to adaptively decrease their distances to the users is essential for the future communication networks. Motivated by this, aerial base stations (ABSs) have recently attracted significant attention due to their applications in wireless networks 
%\cite{Survey_Implementation},
\cite{ Survey_RZhang}. Considering advantages such as high mobility, low cost, and flexible deployement offered by the ABSs \cite{R_Zhang_Survey}, they can be considered as a promising technology to improve the coverage of mobile users in the wireless networks.

%high mobility, low cost, and flexible deployement are among the most important advantages . 
%finding a new architecture to offer a hybrid behavior for the base stations 
%More coverage, supporting more devices, and higher data rate demands are among the most critical concerns of the future mobile networks. 

%Moreover, due to the existing long distances between the ground BSs and mobile users, the coverage can be poor so that the required quality of service (QoS) of the users is not met. 
  %.  Due to high mobility, low cost, and flexible deployement \cite{R_Zhang_Survey}, the UAVs can be considered as a promising technology to improve the coverage of mobile users in the wireless networks.

The conducted research for the ABSs can be categorized into two lines: static ABSs and dynamic ABSs. For the statice ABSs, the objective is to find the optimal placement of the ABS(s) in a way that some criteria such as coverage or sum-rate of the users is maximized \cite{Placement_optimization_RZhang, Alzenad_CL, Rozhina}. %, Mozaffari_Deployment_CL}. 
For instance, in \cite{Placement_optimization_RZhang} the optimal placement of the ABSs is determined to minimize the number of ABSs required to cover ground users, ensuring that each mobile user is in the coverage area of at least one of the ABSs. In \cite{Alzenad_CL} an algorithm is developed to optimize the 3-D deployment of an ABS with the objective of maximizing the number of covered users using the minimum transmit power. 
%In \cite{Mozaffari_Deployment_CL}, using the circle packing theory, optimal 3D deployment of the UAVs is found such that the total coverage area is maximized. 
%The authors in \cite{Rozhina} propose an algorithm based on Q-learning to find the optimal location of a UAV considering mobility of the users. 
However, in all of these work, the location of the ABS is assumed to be fixed. For the dynamic ABSs, however, the main goal is to leverage the mobility of the ABSs. Thus, the trajectory of the ABS(s) is optimized to improve the performance of the mobile users 
 \cite{TWC_joint_traj_multi_UAV, TVT_Offloading_dell_edge, Common_throughput_RZhang}. %TWC_energy_efficient_UAV_RZhang}. 
This performance gain is resulted from the fact that when the distance between an ABS and user decreases, the probability of having a line-of-sight (LoS) link increases, and hence, higher data rates are achievable in comparison to the static case. In \cite{TWC_joint_traj_multi_UAV} based on the block coordinate descent and successive convex approximation techniques, the authors optimize the user association, ABS trajectories, and transmit power of a multi-ABS system with the objective of maximizing the minimum data rate of the users. %In \cite{CL_Power_efficient_traj}, the authors developed an algorithm to find the trajectory of the UAV with the objective of minimizing the total power consumption of the UAV. 
In \cite{TVT_Offloading_dell_edge}, data offloading problem for an ABS-enabled wireless network has been investigated and the trajectory of the ABS obtained with the objective of maximizing the sum-rate of the users served by the ABS, while minimum data requirement has been considered for all users. In \cite{Common_throughput_RZhang} the authors proposed an algorithm to maximize the minimum data rate of ground users by jointly optimizing the ABS trajectory and its power and bandwidth. 
%In \cite{TWC_energy_efficient_UAV_RZhang} the authors investigated the energy efficient UAV communication problem. By deriving a theoretical model on the propulsion energy consumption of the UAVs, they developed algorithms to find the trajectory of the UAV for both rate-maximization and energy-minimization problems. 
It is worth mentioning that 
%although all of these research are focused on optimizing the trajectories of the UAVs, they 
in all of these work, it is assumed that perfect knowledge of the environment is available. 
%channel state information (CSI) . 
This assumption is not practical since the topology of the network can change continuously, and to have perfect knowledge of the environment, a significant amount of information has to be exchanged between the ABSs and the core network, which is not possible.

%. Moreover, this assumption needs  exchange 
%Therefore, the mobile UAVs are prefered over the static UAVs. 
Motivated by this, 
%in this paper 
we propose a distributed algorithm based on Q-learning to optimize the trajectory of the ABSs. In contrast to the existing algorithms, our algorithm does not need perfect knowledge of the environment, and the amount of information exchanged between ABSs and the core network is negligible. The objective of our trajectory design algorithm is to maximize the sum-rate of users served by each ABS. %while the flight time of the UAVs is kept reasonably short to limit the propulsion energy consumption of the UAVs. 
Therefore, %not only the trajectory optimization is essential, but also the optimal power and sub-channel allocation is 
it is of great importance to allocate optimal power and sub-channels to the users to support them with the highest data rates. To solve this complicated problem, we divide it into two sub-problems: trajectory optimization sub-problem (higher level) and joint power and sub-channel allocation sub-problem (lower level). In the higher level, based on the reinforcement learning techniques, a Q-learning problem is formulated to train and update the trajectory of the ABSs using the feedback signal received from the environment. On the other hand, in the lower level, a joint power and sub-channel allocation problem is solved to form the reward function emerging from taking actions by the ABSs in the higher level. %This process repeats until the optimal trajectory is found.

The rest of the paper is organized as follows: Section II describes the system model. The channel model is also discussed in this section. Principals of reinforcement learning is presented in section III. Section IV formulates the trajectory design problem as a Q-learning problem. Our algorithm is also presented in section IV. Simulation results and conclusion are presented in section V and VI, respectively.

\section{System Model}
In this paper, 
%as depicted in Fig. , 
we consider the downlink of a wireless network integrated with multiple ABSs. The set of all ABSs is presented by $\mathcal{J}=\{1,2, \ldots, J\}$, and the set of users associated with the $j$-th ABS is denoted by $\mathcal{K}_j$. Moreover, the ABSs share $\mathcal{N}=\{1,2,\ldots, N\}$ orthogonal sub-channels. We use indices $j$, $k$, and $n$ to represent ABSs, mobile users, and sub-channels. The position of the $j$-th ABS at time $t$ is denoted by ${\bf{\Omega}}_j (t)=(x_j (t), y_j (t),H)$, where $H$ is the altitude of the ABSs which is assumed to be constant throughout their flight time. The initial and final position of the $j$-th ABS are indicated by ${\bf{\Omega}}_j^{0}=(x_j^0,y_j^0,H)$ and ${\bf{\Omega}}_j^{F}=(x_j^F,y_j^F,H)$, respectively. According to this notation, the trajectory of the $j$-th ABS starts from ${\bf{\Omega}}_j^{0}$ and ends to ${\bf{\Omega}}_j^{F}$. 
%We suppose that the initial and final positions of the UAVs are equipped with energy stations so UAVs can charge their batteries once they arrive in these locations.  
%This is crucial for the UAVs 
We assume that the speed of the ABSs is constant during their flights, i.e.,
%\begin{equation}
%\label{velocity_UAVs}
%\sqrt{ \left(V_{j,x} (t)\right)^2 + \left(V_{j,y} (t)\right)^2}
$\rVert V_j (t)\rVert  =V$, $\forall t,j$, 
%\end{equation}
where $V$ is the constant velocity. The distance between ABS $j$ and its final position ${\bf{\Omega}}_j^{F}$ at time $t$ is expressed as
\begin{equation}
\label{Distance_2_Final_position}
D_j (t) \triangleq \Big\rVert {\bf{\Omega}}_{j} (t) - {\bf{\Omega}}_{j}^F \Big\rVert^2, \hspace{0.5 cm} \forall j, t,  %\\
%=\sqrt{ \Big(  x_{j} (t)-x_{j}^F \Big)^2 + \Big(  y_{j} (t)-y_{j}^F \Big)^2  + H^2 },
\end{equation}
and the distance between ABS $j_1$ and ABS $j_2$ at time $t$ is
\begin{equation}
\label{Distance_2pair_UAVs}
D_{j_1,j_2} (t) \triangleq \Big\rVert {\bf{\Omega}}_{j_1} (t) - {\bf{\Omega}}_{j_2} (t) \Big\rVert^2, \hspace{0.5cm} \forall j_1 \neq j_2, \forall t.  %\\
%= \sqrt{ \Big(  x_{j_1} (t)-x_{j_2} (t) \Big)^2 + \Big(  y_{j_1} (t)-y_{j_2} (t) \Big)^2 }.
\end{equation}
To avoid any collision between the ABSs, their trajectories are subject to collision avoidance constraint as follows
\begin{equation}
\label{safety}
D_{j_1 , j_2} (t) > D_{\text{min}}, \hspace{0.3cm} \forall j_1 \neq j_2, \forall t,
\end{equation}
where $D_{\text{min}}$ is the minimum distance between any pair of ABSs that ensures collision avoidance.

%In what follows, the channel model of the air-to-ground links are discussed.
%
%\subsection{Channel Model}
For the air-to-ground links between ABSs and mobile users, we assume that the channel gains are composed of path loss and frequency selective Rayleigh fading. If $h_{jk}^n (t)$ denotes the channel between the $j$-th ABS and user $k$ over the $n$-th sub-channel at time $t$, we have 
\begin{equation}
\label{channel}
h_{jk}^n (t)=\frac{\rho_{jk}^n (t)}{\sqrt{{\text{PL}_{jk}} (t)} }, 
\end{equation}
where $\rho_{jk}^n (t)$ accounts for the small scale fading, and ${\text{PL}}_{jk}^n (t)$ is the average path loss of the link at time $t$. For the path loss, we adopt the model proposed in \cite{Optimal_LAP}. According to this model, both LoS and Non-LoS propagation groups are involved in the average path loss expression. To consider the effects of LoS and Non-LoS links on the average path loss, we need to know their probabilities which depend on the elevation angle between the ABS and the mobile users. Based on \cite{Optimal_LAP}, the probability of having a LoS link between the $j$-th ABS and user $k$ at time $t$ is given by
\begin{equation}
\label{LoS_Pr}
{\text{pr}}_{jk}^{\text{LoS}} (t)=\frac{1}{1+a{\text{ exp}}\left(-b(\theta_{jk}(t)-a)\right)},
\end{equation}
where $\theta_{ik} (t)$ is the elevation angle between ABS $j$ and user $k$ at time $t$, and $a$ and $b$ are constants depending on the environment. The elevation angle is given as $\theta_{ik} (t)=\arctan(\frac{H}{L(t)})$, where $H$ and $L(t)$ are the altitude and the horizontal distance between the $j$-th ABS and user $k$ at time $t$, respectively. Based on the probability given in (\ref{LoS_Pr}), the average path loss between ABS $j$ and user $k$, ${\text{PL}}_{jk} (t)$ can be formulated as 
\begin{equation}
\label{avg_PL}
{\text{PL}}_{jk}(t)= {\text{pr}}_{jk}^{\text{LoS}}(t) {\text{PL}}^{\text{LoS}}(t) + (1-{\text{pr}}_{jk}^{\text{LoS}}(t)) {\text{PL}}^{\text{N-LoS}}(t),
\end{equation}
where ${\text{PL}}^{\text{LoS}}(t)$ and ${\text{PL}}^{\text{N-LoS}}(t)$ are the free space path losses corresponding to the LoS and Non-LoS links, respectively. The free space path loss corresponding to the LoS is expressed as
%\begin{equation*}
${\text{PL}}^{\text{LoS}}(t)=\left(\frac{4\pi f_c d_{jk}(t)}{c}  \right)^2 \times \eta^{\text{LoS}}$,
%\end{equation*}
where $d_{jk}(t)$ represents the distance between the $j$-th ABS and user $k$ at time $t$, $f_c$ is the carrier frequency, $c$ is the speed of light, and $\eta^{\text{LoS}}$ is the environment-dependant additional loss due to LoS link. In a similar way, the free space path loss corresponding to the Non-LoS is described as
%\begin{equation*}
${\text{PL}}^{\text{N-LoS}}(t)=\left(\frac{4\pi f_c d_{jk}(t)}{c}  \right)^2 \times \eta^{\text{N-LoS}}$, 
%\end{equation*}
where $\eta^{\text{N-LoS}}$ is the additional loss due to the Non-LoS link. For the small scale fading, without loss of generality we assume that for $\forall j,k,n$,  $\rho_{jk}^n (t)$ are independent and identically distributed (i.i.d) random variables with $\mathbb{E} \{ |\rho_{jk}^n|^2 \}=1$. Using the described model, the channel gain between the $j$-th ABS and user $k$ at time $t$ is given by 
\begin{equation}
\label{channel_gain}
g_{jk}^n (t)\triangleq \left| h_{jk}^n (t)\right|^2.
\end{equation}

%\subsection{Communication Model} 
As discussed earlier, in this paper, we are to find trajectories of the ABSs with the objective of maximizing sum-rate of the users associated with each ABS. %This has to be done in a way that the flight time of the UAVs are kept reasonably short. This is due to the fact that the propulsion power consumption of the UAVs is proportionally increasing with the flight time. 
As a result, in addition to the optimal trajectory design, optimal power and sub-channel allocation is also necessary to support the users with the highest data rates. If the transmit power of the $j$-th ABS on the $n$-th sub-channel and time $t$ is denoted by $P_j^n (t)$, according to Shannon formula, the instanteneous rate of user $k$ served by the $j$-th ABS at time $t$ on sub-channel $n$ is given by
\begin{equation}
\label{rate}
r_{jk}^n (t) = \log \left( 1+ \frac{P_{j}^n (t) g_{jk}^n (t)}{I_{jk}^n(t)+\sigma^2} \right),
\end{equation}
%where $g_{jk}^n (t)$ is %the channel gain between the $j$-th UAV-BS and user $k$ on the $n$-th sub-channel and time $t$, defined in (\ref{channel_gain}),
 where $\sigma^2$ is the thermal noise at the receiver of the $k$-th mobile user and $I_{jk}^n(t)$ is the total interference on the $n$-th sub-channel and time $t$ arising from other ABSs and ground base station (GBS) to the receiver of the $k$-th user. The interference $I_{jk}^n(t)$ is expressed as
%We transmit . The received signal to interference plus noise ratio (SINR) on sub-channel $n$ and time $t$ for the mobile user $k$ served by the $j$-th UAV-BS is given by
%\begin{equation}
%\label{SINR}
%\frac{P_{j}^n (t) h_{jk}^n (t)}{I_{jk}^n(t)+\sigma^2}
%\end{equation}
\begin{equation}
\label{Interference}
I_{jk}^n(t)= \displaystyle \sum_{j'=1, j' \neq j}^{J} P_{j'}^n (t) g_{j'k}^n (t) + P_{0}^n (t) g_{0k}^n (t),
\end{equation}
where the first summation is the total interference from the other ABSs and the last term accounts for the interference arising from the GBS. Accordingly, $P_{0}^n (t)$ is the transmit power of GBS at time $t$ over the $n$-th sub-channel, and $g_{0k}^n (t)$ is the channel gain from the GBS to user $k$ over sub-channel $n$ and time $t$.
Moreover, we define the sub-channel assignment indicator as
\begin{equation}
\label{sub_channel_indicator}
a_{jk}^n (t) \in \{0,1\}, \forall j,k,n,t,
\end{equation}
where $a_{jk}^n (t)=1$ indicates that at time $t$, sub-channel $n$ is assigned to the user $k$ by its associating ABS $j$. Otherwise the value of $a_{jk}^n (t)$ is zero. 

\section{Reinforcement Learning Fundamentals}
In this section, we briefly explain the reinforcement learning principals. Then, we formulate our trajectory design problem as a Q-learning problem which can be solved efficiently.
 
\begin{figure}[t]
	\centering
	\includegraphics[width=2.0in,keepaspectratio]{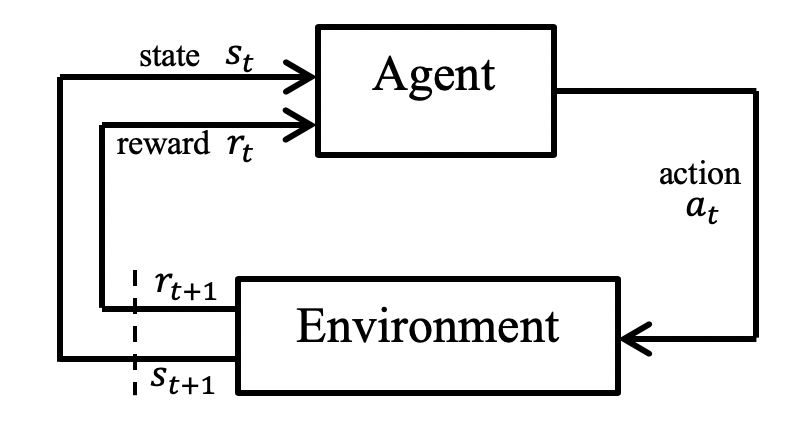}
	\vspace{-0.3em}
	\caption{Interaction between agent and environment.}
	\label{fig_6}
	\vspace{-1.5em}
\end{figure}

In reinforcement learning, the agent interacts with the environment at each of a sequence of discrete time steps denoted by $t=0,1,2,\ldots$. If $s_t$ and $\mathcal{S}$ denote the state occupied by the agent at time $t$ and the set of all possible states, respectively, based on the current state $s_t$, the agent takes an action $a_t \in \mathcal{A}(s_t)$, where $\mathcal{A}(s_t)$ is the set of actions available at state $s_t$. One time step later, agent receives reward $r_{t+1}$ and goes to a new state $s_{t+1}$. Fig. \ref{fig_6} presents the interaction between agent and environment. 

The agent at each time step implements a policy $\pi (a|s)$ which is the probability that $a_t=a$ if $s_t=s$, i.e.,
\begin{equation}
\label{Policy}
\pi (a|s)=\text{Pr} \left( a_t=a | s_t=s \right).
\end{equation}
The implemented policy is modified over time based on the previous experience. Reinforcement learning describes how this policy change is made as a result of agent's experience \cite{Sutton}. The goal of an agent is not maximizing its immediate reward, but the expected sum of discounted reward it receives over the long run is maximized. For this purpose, we define a return function $R_t$ as 
\begin{equation}
\label{Return}
R_t \triangleq \sum_{k=0}^{T-1} \gamma^k r_{t+k+1},
\end{equation}
where $T$ and $\gamma$ are the final time step and the discount rate $0 \leq \gamma < 1$, respectively. Based on (\ref{Policy}) and (\ref{Return}), we can define a function which represents the value of taking action $a$ in state $s$ under policy $\pi$, denoted by $Q_{\pi}(s,a)$ as
\begin{equation}
\label{Q_function}
Q_{\pi} (s,a)= \mathbb{E}_{\pi} \Big\{ R_{t} \Big| s_t=s, a_t=a  \Big\}.
\end{equation}
In other words, $Q_{\pi}(s,a)$ is the expected return starting from $s$, taking action $a$, and following policy $\pi$. This function is called action-value function or simply Q-function. Using the definition of return in (\ref{Return}), the action-value function can be decomposed into immediate reward plus discounted Q-function of successor state as
\begin{equation}
\label{Q_function_decomposed}
Q_{\pi} (s,a)= \mathbb{E}_{\pi} \Big\{ r_{t+1} + \gamma Q_{\pi} (s_{t+1},a_{t+1}) \Big| s_t=s, a_t=a  \Big\}.
\end{equation} 
The optimal Q-function is expressed as
\begin{equation}
\label{optimal_Q}
Q^{\star} (s,a)= \max_{\pi} Q_{\pi} (s,a),
\end{equation}
and the optimal policy can be obtained by maximizing over $Q^{\star} (s,a)$ according to
\begin{equation}
\label{optimal_Policy}
\pi^{\star}(a|s) = 
\begin{cases}
1 &\quad\text{if } a=\arg \max_{a} Q^{\star} (s,a)\\
0 &\quad\text{Otherwise}\\
\end{cases}.
\end{equation}
Moreover, the optimal Q-function has to satisfy the Bellman optimality equation \cite{Sutton} for the Q-function as
\begin{equation}
\label{Bellman_Optimality_Q}
Q^{\star} (s,a) \hspace{-0.05cm} = \hspace{-0.05cm}\mathbb{E}_{\pi} \Big\{ \hspace{-0.05cm} r_{t+1} + \gamma \max_{a'}  Q^{\star} (s_{t+1},a') \Big| s_t \hspace{-0.05cm} = \hspace{-0.05cm} s, a_t \hspace{-0.05cm} = \hspace{-0.05cm} a   \hspace{-0.05cm} \Big\}.
\end{equation}  
It is worth mentioning here that the Bellman optimality equation is non-linear and in general does not have any closed form solution. Alternatively, to solve (\ref{Bellman_Optimality_Q}) we have to use an iterative method. Q-learning is a well-known method for finding $Q^{\star} (s,a)$ in a recursive manner \cite{Watkins}. According to this method, the learned action-value function is updated as 
\begin{multline}
\label{Q_learning}
Q(s_t,a_t) \longleftarrow Q(s_t,a_t) + \alpha \Big[ r_{t+1} + \gamma \max_{a} Q(s_{t+1},a) \\
- Q(s_t,a_t) \Big], 
\end{multline}
where $\alpha \in [0,1]$ is the learning rate. The most important advantage of this method is that the learned Q-function derived by (\ref{Q_learning}) directly approximates the optimal action-value function, $Q^{\star} (s,a)$, independent of the policy followed by the agent \cite{Sutton}.  

\section{Trajectory optimization as a Q-learning problem}
In this section, the trajectory optimization is formulated as a Q-learning problem. To find the trajectories of the ABSs with the objective of maximizing sum-rate of the users associated with each ABS, we divide the problem into two sub-problems: trajectory optimization sub-problem and joint power and sub-channel allocation sub-problem. In other words, a two stage decision making procedure is considered to deal with the original complicated problem. In the first stage, using the Q-learning technique, ABSs take one of the available actions to obtain their trajectories. Upon any change in the position of the ABSs, in the second stage, a new optimization problem is solved to allocate optimal power and sub-channels to the mobile users. The corresponding sum-rate contributes to the reward received from taking the action in the first stage. In what follows, we define the agents, states, actions, and reward associated to the Q-learning framework:
%\begin{itemize}
	%\item 
	\\
	{\bf{Agent}} $j$: ABS $j$, $1 \leq j \leq J$.\\
	%\item 
	{\bf{State}} $s_t^{(j)}$: Position of ABS $j$ at time $t$. In other words, $s_t^{(j)}=\{x_{j}(t),y_{j}(t),H\}$. It is worth mentioning that in general, the position of each ABS can be a continuous function. To restrict the number of possible states, we place a grid with limited number of squares on it to discretize the area. The center of each square represents that state. For instance, if the area of interest is  rectangular and is presented by ${\text{x}}_{\text{min}} \leq x \leq {\text{x}}_{\text{max}}$ and ${\text{y}}_{\text{min}} \leq y \leq {\text{y}}_{\text{max}}$, we convert this continuous area into $M^2$ states by spliting both $[{\text{x}}_{\text{min}},{\text{x}}_{\text{max}}]$ and $[{\text{y}}_{\text{min}},{\text{y}}_{\text{max}}]$ intervals into $M$ slots. In the resulting grid, the coordinates of the center of the square located at the $k_1$-th slot in the ${\text{x}}$ axis and the $k_2$-th slot in the ${\text{y}}$ slot are 
	%\begin{subequations}
	%\begin{align}
	${\text{x}}_{k_1}={\text{x}}_{\text{min}} + \frac{({\text{x}}_{\text{max}}-{\text{x}}_{\text{min}}) }{M}(k_1 -1)$ and %, \nonumber \\
	${\text{y}}_{k_2}={\text{y}}_{\text{min}} + \frac{({\text{y}}_{\text{max}}-{\text{y}}_{\text{min}}) }{M}(k_2 -1)$, respectively. %. \nonumber
	%\end{align}
	%\end{subequations}
	As a result, the number of available states in our problem is limited to $M^2$.\\
	%\item 
	{\bf{Action}} $a_t^{(j)}$: Available actions at each state are the movement in four directions: left, right, forward, and backward. \\
	%\item 
	{\bf{Reward}} $r_t^{(j)}$: The reward function for the $j$-th agent is defined as follows
	\begin{equation}
	\label{Reward_definition}
	r_t^{(j)}=\beta_1 F^{(j)}_1 (t) - \beta_2 F^{(j)}_2 (t) - \beta_3 F^{(j)}_3 (t),
	\end{equation}
	where $F_1^{(j)}$ reflects the sum-rate of the users served by ABS $j$, $F_2^{(j)}$ motivates the ABS to complete its flight in a reasonably short time, and $F_3^{(j)}$ is an activation function required for the safety of the ABSs. In (\ref{Reward_definition}),  $\beta_1$, $\beta_2$, and $\beta_3$ are parameters of the reward function. In what follows we introduce functions $F_1^{(j)}$, $F_2^{(j)}$, and $F_3^{(j)}$ in more details and explain the rationale behind this selection.
	
	As discussed earlier, Q-learning aims to maximize the sum-rate of the users while the time to complete the flight is reasonably short. Moreover, for safety purposes, the distance between any pair of ABSs must be greater than some predefined threshold. Based on these concerns, we design the reward function. Function $F_1^{j}(t)$ is defined to be the sum-rate of mobile users currently served by the ABS $j$. This function is expressed as
	\begin{equation}
	\label{F1}
	F_1^{(j)}(t) \hspace{-0.1cm}= \hspace{-0.1cm} \sum_{k=1}^{K} \sum_{n=1}^{N} \hspace{-0.07cm} a_{jk}^{\star n} (t) \log \hspace{-0.03cm} \left(\hspace{-0.05cm}1+\hspace{-0.05cm} \frac{P_{j}^{\star n} (t) g_{jk}^n (t)}{I_{jk}^n(t)+\sigma^2}\right),
	\end{equation}
	where $P_{j}^{\star n} (t)$ and $a_{jk}^{\star n} (t)$ are the optimal power and sub-channel allocation at time $t$, respectively. To find these values, the $j$-th ABS  solves the following optimization problem at time $t$
\begin{subequations}
	\begin{align}
	\label{P_Ch_allocation}
	\max_{P, a} \hspace{0.5cm}& \sum_{k=1}^{K} \sum_{n=1}^{N} a_{jk}^{n} (t) \log \left(1+\frac{P_{j}^n (t) g_{jk}^n (t)}{I_{jk}^n(t)+\sigma^2}\right) \\
	%& \text{s.t. } \nonumber \\
	\text{s.t. } \hspace{0.2cm} & \begin{matrix} \label{Sum_power}  \displaystyle \sum_{k=1}^{K} \displaystyle \sum_{n=1}^{N} a_{jk}^{n} (t) P_j^n (t) \leq P_{\text{max}}, \end{matrix} \\
	& \begin{matrix}\label{sub_ch_add2one} \displaystyle \sum_{k=1}^{K} a_{jk}^{n} (t) \leq 1,  \hspace{0.5cm}\forall n, \end{matrix} \\
	& \begin{matrix} \label{Binary_sub_ch} a_{jk}^{n} (t) \in \{0,1\}, \hspace{0.5cm} \forall k,n, \end{matrix}
	\end{align}
\end{subequations}
where constraint (\ref{Sum_power}) ensures that the transmit power of each ABS is limited to $P_{\text{max}}$ and (\ref{sub_ch_add2one}) guarantees that at each time $t$, each sub-channel is assigned to at most one mobile user.  
The optimization problem of (\ref{P_Ch_allocation}) is a non-convex mixed integer problem due to constraint (\ref{Binary_sub_ch}). To make it tractable, we first relax the binary variables $a_{jk}^{n} (t)$ to be continuous in the interval of $[0,1]$. Moreover, we denote the actual power allocated to the $k$-th mobile user served by the $j$-th ABS on the $n$-th sub-channel and time $t$ with $\tilde{P}_{jk}^n (t)\triangleq a_{jk}^n (t) P_{j}^n (t)$. As a result, the new optimization problem is given by 
\begin{subequations}
	\begin{align}
	\label{P_Ch_allocation_relaxed}
	\max_{\tilde{P}, a} \hspace{0.0cm}& \sum_{k=1}^{K} \hspace{-0.08cm} \sum_{n=1}^{N} \hspace{-0.08cm} a_{jk}^{n} (t) \log \hspace{-0.08cm} \bigg(\hspace{-0.08cm} 1 \hspace{-0.08cm}+ \hspace{-0.08cm}\frac{\tilde{P}_{j}^n (t) g_{jk}^n (t)}{ a_{jk}^n (t) \big( I_{jk}^n(t)+\sigma^2 \big)}\hspace{-0.07cm} \bigg) \\
	%& \text{s.t. } \nonumber \\
	\text{s.t. } \hspace{0.2cm} & \begin{matrix} \label{Sum_power_relaxed}  \displaystyle \sum_{k=1}^{K} \displaystyle \sum_{n=1}^{N} \tilde{P}_{jk}^n (t) \leq P_{\text{max}}, \end{matrix} \\
	& \begin{matrix} \label{Binary_sub_ch_relaxed} 0 \leq a_{jk}^{n} (t) \leq 1, \hspace{0.5cm} \forall k,n, \end{matrix}\\ \hspace{-0.1cm}
	& (\ref{sub_ch_add2one}). \nonumber
	\end{align}
\end{subequations}

It can be shown that the function $f(x,y)=x\log (1+\frac{y}{x}h)$, for a constant $h$, is concave over $x\geq 0$ and $y\geq 0$ since its Hessian matrix is negative semi-definite. Therefore, problem of (\ref{P_Ch_allocation_relaxed}) is a convex problem. To solve (\ref{P_Ch_allocation_relaxed}), we form its Lagrangian function as follows
\begin{multline}
\label{Lagrangian} 
\mathcal{L}\{ {\bf{P}}, {\bf{a}}, \lambda, {\bf{\mu}} \} \hspace{-0.05cm}=  \hspace{-0.1cm} \sum_{k=1}^{K} \hspace{-0.1cm} \sum_{n=1}^{N} \hspace{-0.05cm} a_{jk}^{n} (t) \log \left( \hspace{-0.08cm} 1 \hspace{-0.08cm}+\frac{P_{j}^n (t) g_{jk}^n (t)}{I_{jk}^n(t)+\sigma^2}\right) \\ 
+ \lambda \hspace{-0.1cm} \left( \hspace{-0.1cm} P_{\text{max}} \hspace{-0.07cm}- \hspace{-0.08cm} \displaystyle \sum_{k=1}^{K} \hspace{-0.07cm} \displaystyle \sum_{n=1}^{N} \hspace{-0.08cm} a_{jk}^{n} (t) P_j^n \hspace{-0.02cm} (t) \hspace{-0.1cm} \right) \hspace{-0.08cm} + \hspace{-0.07cm} \displaystyle \sum_{n=1}^{N} \hspace{-0.04cm} \mu_n \hspace{-0.1cm} \left( \hspace{-0.1cm} 1 \hspace{-0.05cm}- \hspace{-0.08cm} \displaystyle \sum_{k=1}^{K} \hspace{-0.08cm} a_{jk}^{n} (t) \hspace{-0.1cm} \right), 
\end{multline}
where $\lambda$ and $\mu_n$ are the dual multipliers associated with constraints (\ref{Sum_power}) and (\ref{sub_ch_add2one}), respectively. The Lagrangian dual function is defined as 
\begin{equation*}
\label{Lagrangian_Dual}
g(\lambda, {\bf{\mu}})= \max_{{\tilde{\bf{p}}}, {\bf{a}}} \mathcal{L}\{ {\bf{P}}, {\bf{a}}, \lambda, {\bf{\mu}} \},
\end{equation*}
and the dual problem is 
\begin{equation}
	\begin{aligned}
	\label{Dual_Problem}
	 &\min_{\lambda, {\bf{\mu}}} \hspace{0.3cm} g(\lambda, {\bf{\mu}}) \hspace{0.5cm} \text{s.t. }  \begin{matrix}  \lambda, {\bf{\mu}} \geq 0.  \end{matrix} % \\
	 %&\text{s.t. }  \begin{matrix}  \lambda, {\bf{\mu}} \geq 0,  \end{matrix}  \\
	\end{aligned}
\end{equation}

According to the KKT conditions \cite{convexOpt}, the optimal solution of (\ref{P_Ch_allocation_relaxed}) must sastisfy $\frac{\partial \mathcal{L}}{\partial {\tilde{P}}_{jk}^n }=0$ and $\frac{\partial \mathcal{L}}{\partial a_{jk}^n }=0$, As a result, we obtain
\begin{equation}
\label{Optimal_power_allocation}
P^{\star n}_{j} (t)=\frac{{\tilde{P}}_{jk}^{\star n} (t)}{a_{jk}^n (t)} = \left[ \frac{1}{\ln 2 \hspace{0.1cm}\lambda} - \frac{  I_{jk}^n  (t)+ \sigma^2 }{h_{jk}^n (t)}\right] ^+,
\end{equation}
and
\begin{equation}
\label{Optimal_sub_channel}
a_{jk^{\star}}^n (t)= 1 \Big|_{k^{\star}=\arg \max_{k} \Psi_{jk}^n (t)},
\end{equation}
where
\begin{multline*}
\label{Phi}
\Psi_{jk}^n (t)= \log \Big( 1+\frac{P^{\star n}_{j} (t) h_{jk}^n (t)}{ I_{jk}^n (t)+ \sigma^2} \Big) - \\ \frac{1}{\ln 2} \frac{P^{\star n}_{j} (t) h_{jk}^n (t) }{P^{\star n}_{j} (t) h_{jk}^n (t)+  I_{jk}^n (t)+\sigma^2  } .
\end{multline*}
In other words, sub-channel $n$ is assigned to user $k$ with the largest value of $\Psi_{jk}^n (t)$.
The dual variable $\lambda$ can be updated with the sub-gradient method according to
\begin{equation}
\label{lambda_update}
\lambda (l+1) \hspace{-0.1cm}= \hspace{-0.1cm} \left[ \lambda (l) - \alpha (l) \Big(\hspace{-0.05cm} P_{\text{max}} \hspace{-0.05cm}- \hspace{-0.05cm}\displaystyle \sum_{k=1}^{K} \hspace{-0.05cm} \displaystyle \sum_{n=1}^{N} \hspace{-0.05cm} \tilde{P}_{jk}^n (t)  \Big)   \right]^+,
\end{equation}
where $\alpha (l)$ is the step size in the $l$-th iteration which has to satisfy %the following constraints
%\begin{equation}
%\label{step_size_constraints}
%\begin{matrix}
$\displaystyle \sum_{l=1}^{\infty} \alpha (l) = \infty$, and $\lim\limits_{l \to \infty} \alpha (l) =0$.
%\end{matrix}
%\end{equation} 
Using (\ref{Optimal_power_allocation}) and (\ref{Optimal_sub_channel}), the sum-rate of the users associated with the $j$-th ABS is derived by (\ref{F1}).

	\begin{algorithm}[t]
	\footnotesize
	%\scriptsize
	\caption{Distributed Q-Learning for the ABS trajectory design}
	\label{Q_learning_ALg}
	\begin{algorithmic}[1]
		\State For all ABSs, initialize their Q-function $Q_j(s,a)$ for all $s \in \mathcal{S}$, $\forall a \in \mathcal{A} (s)$ and $\forall j \in \mathcal{J}$, arbitrarily.
		\State Set $Q_j(\text{Terminal state},.)=0$ for all ABSs.
		\For {episode=1 to max episode}
		\State Initialize $s^{(j)}_0$ based on the initial location of each ABS $j$.
		%\Repeat For each step of episode
		\For {each step of episode (time $t$)}
		\For {each ABS}
		\If {$rand(.) < \epsilon$}
		\State select action randomly (exploration)
		\Else
		%\State Choose $a$ from $s$ using policy derived from $Q$ (e.g. $\epsilon$-greedy)
		\State Choose action $a^{(j)}_t=\arg \max_{a'} Q_j(s^{(j)}_t,a')$ (exploitation)
		\EndIf
		\State Take action $a^{(j)}_t$, 
		\State Receive the immediate reward, $r_{t+1}^{(j)}$ according to (\ref{Reward_definition}) 
		\State Observe the new state $s^{(j)}_{t+1}$
		\State Update $Q$-function for ABS $j$ as
		\State $Q_j(s^{(j)}_t,a^{(j)}_t) \longleftarrow Q_j(s^{(j)}_t,a^{(j)}_t) + \alpha \Big[r^{(j)}_{t+1} + \gamma \max_{a} Q_j(s_{t+1}^{(j)},a) -Q_j(s^{(j)}_t,a^{(j)}_t)\Big] $
		%\State $s \longleftarrow s_{t+1}^{(j)}$ 
		%\Until $s$ is terminal state.
		\EndFor
		\EndFor
		\EndFor
	\end{algorithmic}
\end{algorithm}

In addition to the sum-rate function $F^{(j)}_1 (t)$, we need another function $F^{(j)}_2 (t)$ to motivate the ABS to reach the terminal state (final position) at the end of its flight time. This prevents ABSs from staying at a specific point, and forces the ABSs to reach their destinations. We define 
\begin{equation}
\label{F2}
F^{(j)}_2 (t)= D_{j} (t),
\end{equation}
%$F^{(j)}_2 (t)= D_{j} (t)$, 
where $D_{j} (t)$ is the distance between ABS $j$ and its final position at time $t$ and is defined in (\ref{Distance_2_Final_position}). Moreover, we need another function $F^{(j)}_3 (t)$ to act like an activation function returning a value when the distance between ABS $j$ to any other ABS is less than the threshold distance required for the collision avoidance. In other words,
\begin{equation}
\label{F3}
F^{(j)}_3 (t)= \begin{cases}
1&\quad D_{j,j'} (t) < D_{\text{min}}, \forall j' \neq j\\
0 &\quad\text{Otherwise}\\
\end{cases}.
\end{equation}
Based on (\ref{F1}), (\ref{F2}), and (\ref{F3}), the reward function is expressed as (\ref{Reward_definition}).  
%\end{itemize}
Algorithm \ref{Q_learning_ALg} shows our distributed Q-learning algorithm for the ABS trajectory design.

%is the distance between the UAV $j$ and its terminal state. In other words, this term is added to motivate the UAV to go to the terminal state. Otherwise,  maximized.

\section{Simulation Results}
In this section, numerical results are presented to show the performance of the proposed algorithm. We consider a $3$km $\times$ $3$km area which using a rectangular grid is divided into $M^2=900$ states. Users are randomly distributed in the area. In our simulations we assume a dual-ABS system. The number of sub-channels and the carrier frequency are $N=8$ and $f_c=2$GHz, respectively. The total number of users in our simulations is 20 in which half of them are assigned to each one of the ABSs. The altitude of the ABSs is assumed $H=100$m. Other parameters are as follows: $(a,b)=(5,0.5)$, $\eta^{\text{LoS}}=1$, $\eta^{\text{N-LoS}}=20$,  $\gamma=0.9$, $\epsilon=0.1$, $(\beta_1,\beta_2,\beta_3)=(10,0.25,1000)$
, $V=10$m/s, $P_{\text{max}}=200$mW, %$\sigma^2 = $, 
$D_{\text{min}}=5$m, $[{\bf{x}}_{min},{\bf{x}}_{max}]=[0,3000]$, and $[{\bf{y}}_{min},{\bf{y}}_{max}]=[0,3000]$.

%
%
%it has a good performance and can train the agents of our network in an efficient way.. 

\begin{figure} [t]
	\centering
	\includegraphics[width=3.3in,keepaspectratio]{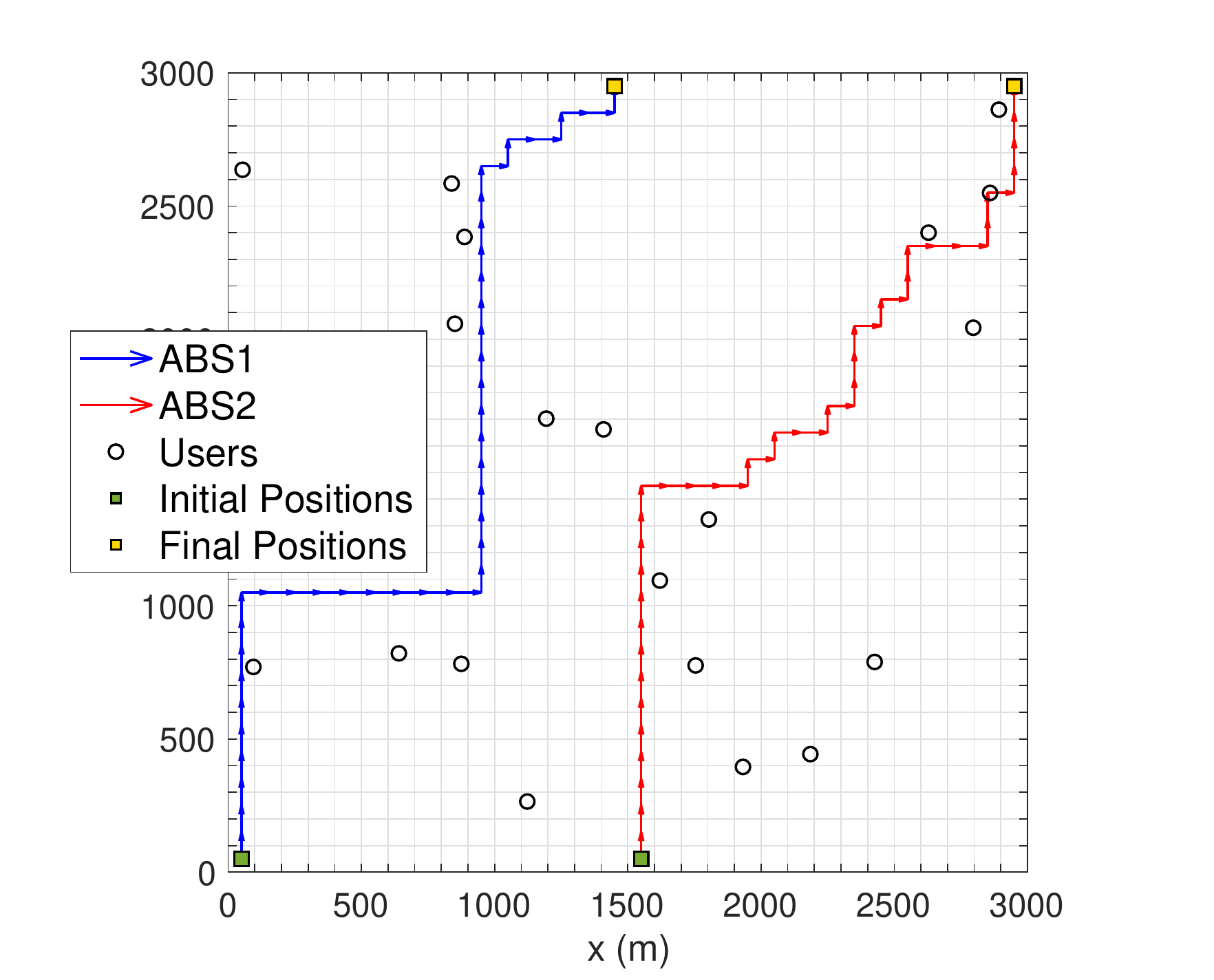}
	\vspace{-1.8em}
	\caption{Final trajectory of the ABSs.}
	\label{Fig_Final_trajectory}
	\vspace{-1.8em}
\end{figure}

Fig. \ref{Fig_Final_trajectory} presents the final trajectory of the ABSs. This figure shows that although Q-learning is a model-free method, it has ability to make the agents aware of the environment. In other words, using the feedback signals which are the rewards signals, the ABSs can be trained to find their trajectories based on the topology of the network. This figure also shows that in addition to moving from the initial position toward the final position, the ABSs repeatedly decrease their distances to their associated users. This is essential for the ABSs since by reducing the distance to a user, the link quality between the ABS and the aformentioned user is improved. As a result, the data rate of the user increases. Moreover, it is observed that the ABSs eventually reach to their final positions either to get prepared for the next flight time or to recharge their batteries. This is done so that the distance between the ABSs at any time during their flight remains more than $D_{\text{min}}$.

\begin{figure}[t]
	\centering
	\includegraphics[width=3.2in,keepaspectratio]{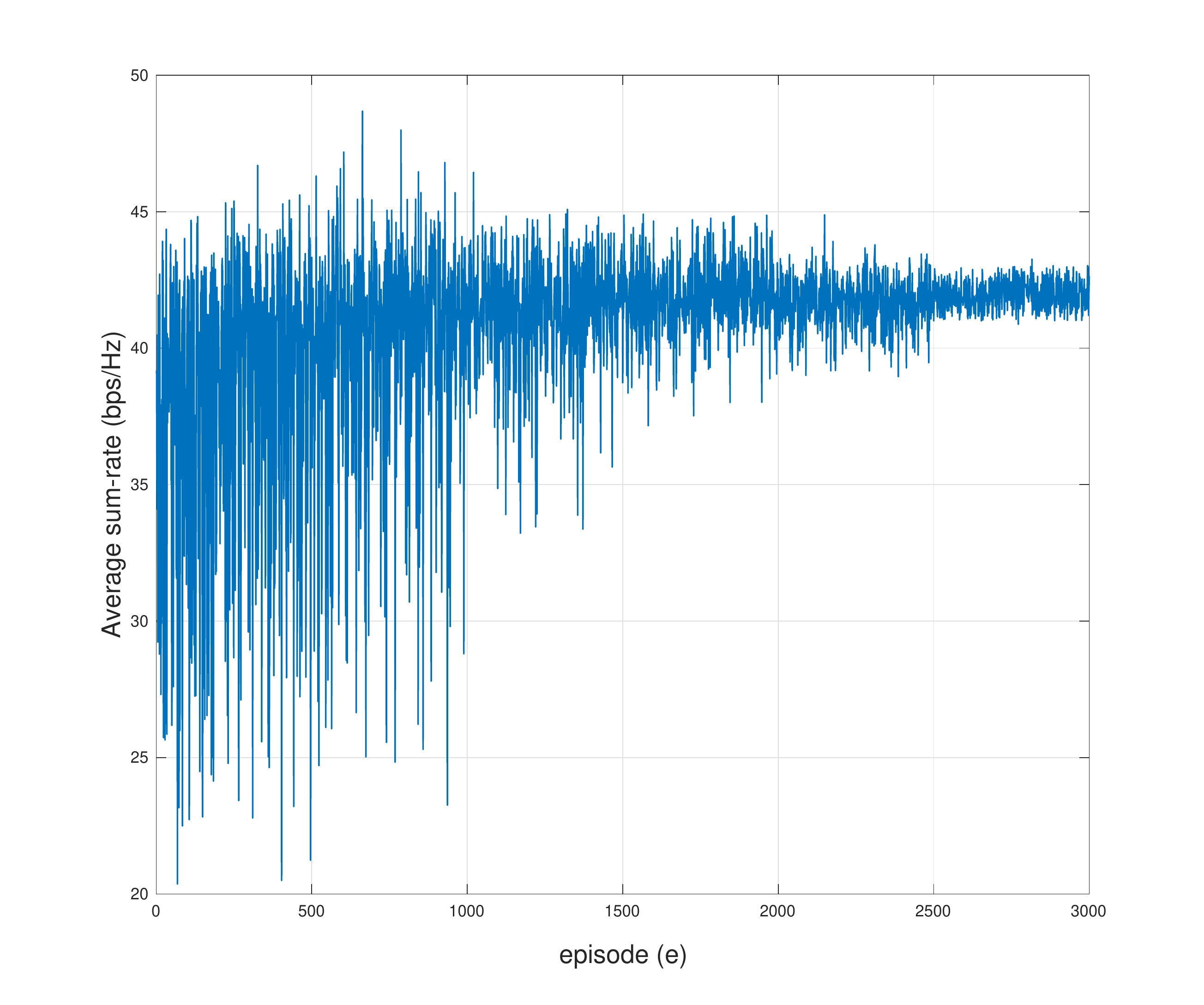}
	\vspace{-1.8em}
	\caption{Convergence of the average sum-rate of the users}
	\label{Fig_Convergence}
	\vspace{-1.8em}
\end{figure}

Fig. \ref{Fig_Convergence} shows convergence of the average sum-rate of the users. Index $e$ indicates the episode number. As can be observed, at the begining of the learning process, the achievable sum-rate fluctuates widely. However, after a sufficient number of episodes, this fluctuation becomes negligible and the sum-rate converges to its optimal value. This is due to the fact that the agents (ABSs) are trained based on the feedback signals they receive. These feedback signals carry required information from the environment. Therefore, After a sufficient number of episodes, the ABSs will have a decent knowledge of the topology of the network. This improves the performance of the ABSs in terms of their trajectory and their achievable data rates.

\section{Conclusion}
In this paper, we studied the trajectory optimization problem for a wireless network integrated with multiple ABSs. The objective is to maximize the total data rate of users served by each ABS. As a result, in addition to the trajectory optimization, it is of great importance to allocate optimal power and sub-channels to the users to support them with the highest possible data rates. To address all, we divide the problem into two sub-problems: trajectory optimization sub-problem and joint power and sub-channel allocation sub-problem. For the trajectory optimization sub-problem, a reinforcement learning problem has been formulated while for the later one an optimization problem has been solved. The simulation results show that although Q-learning is a model-free reinforcement learning method, it effectively trains the ABSs to optimize their trajectories.
\vspace{-0.4em}
\bibliographystyle{IEEEtran}
\bibliography{Citations}

\end{document}